\def\etal{{\it et al.~\/}}

\def\ltsima{$\; \buildrel < \over \sim \;$}
\def\simlt{\lower.5ex\hbox{\ltsima}}
\def\gtsima{$\; \buildrel > \over \sim \;$}
\def\simgt{\lower.5ex\hbox{\gtsima}}

\documentstyle[12pt,aasms4,epsfig]{article}
\tighten
\singlespace

\lefthead{Aut1 ,Aut2, Aut3 ....}
\righthead{ TitleTitle }

\begin{document}

\title{ Smoothed particle hydrodynamics for galaxy formation
simulations: improved treatments of multiphase gas,
of star formation and of supernovae feedback }

\author{S. Marri$^{1}$, S. D. M. White$^{2}$}
\affil{
	$^1$~Max-Planck-Institut fuer Astrophysik, \\
	85741 Garching bei Munchen, Germany \\
	E--mail: marri@mpa-garching.mpg.de \\
        $^2$~Max-Planck-Institut fuer Astrophysik, \\
        85741 Garching bei Munchen, Germany \\
        E--mail: swhite@mpa-garching.mpg.de \\
      }

\begin{abstract}
We investigate a new implementation of the Smoothed Particle
Hydrodynamics technique (SPH) designed to improve the realism with
which galaxy formation can be simulated. In situations where cooling
leads to the coexistence of phases of very different density and
temperature, our method substantially reduces artificial overcooling 
near phase boundaries, prevents the exclusion of hot gas from the 
vicinity of cold ``clouds'', and allows relative motion of the two
phases at each point. We demonstrate the numerical stability 
of our scheme in the presence of extremely steep density and temperature
gradients, as well as in strong accretion shocks and cooling flows.
In addition, we present new implementations of star formation and 
feedback which simulate the effect of 
energy injection into multiphase gas more successfully than previous
schemes.  Our feedback recipes deposit thermal energy separately in
cold dense gas and hot diffuse gas, and can explicitly reinject
cold gas into the hot phase. They make it possible to 
damp star formation effectively, to reheat cold gas, and to drive 
outflows into the galaxy halo and beyond. We show feedback effects
to be strongest in small mass objects where much  
of the gas can be expelled. After idealised tests, we carry out a 
first low resolution study of galaxy formation in a 
$\Lambda$CDM universe. Feedback results in substantial and mass-dependent
reductions in the total baryonic mass gathered onto the final object
as well as in significant modulation of the star formation history.

\end{abstract}


\section{Introduction.}
A detailed understanding of galaxy formation in cold dark matter universes 
remains a primary goal of modern astrophysics. Whereas on large
scales the clustering of matter is determined almost solely by  
gravitational forces, a large number of other physical processes 
contribute to the dynamics on the scales relevant to galaxy
formation. In order to gain insight into this problem, it is 
important to develop numerical methods which can reliably represent 
these physical processes, many of which occur on scales too small 
to be resolved by the simulations. The aim of this work is to 
describe a set of numerical tools that can be used to simulate
galaxy formation within popular CDM models.

Smoothed Particle Hydrodynamics or SPH (Gingold \& Moneghan 1977;
Lucy 1977) is a particle-based technique for solving gas-dynamics which is
often applied to astrophysical problems. This scheme is fundamentally 
Lagrangian, it can be easily combined with gravity solvers that use tree
structures and it lends itself readily to the wide range of densities 
in galaxy formation problems. However, standard implementations of SPH 
have limited ability to resolve steep density gradients, and a number 
of numerical problems occur when particles are close to
a region of very different density. These arise 
because the usual formulation of SPH assumes that the density gradient 
across the smoothing kernel of each particle is small. This is not
true in many situations in which SPH is commonly used. As a result,
low mass clumps of dense gas artificially ``evaporate'', hot diffuse
gas is prevented from coexisting with dense ``clouds'' and radiative 
cooling is artificially enhanced in the diffuse gas which lies near
such clouds. In most current implementations of SPH the nett result of
these effects is excessive accretion onto the cold phase (Pearce \etal 
1999; Ritchie \& Thomas 2000; Springel \& Hernquist 2002a). 

The method we propose to overcome these
limitations is based on a new neighbour search that considers the 
thermodynamic properties of particles. Our scheme evaluates the
appropriate local
density more correctly than previous schemes for particles near 
a phase boundary. At the same time, it performs well in situations 
where the phases are far from pressure equilibrium, for example 
in shocks. The method will be outlined in Sec. 2.1.  In Sec 3.2, 
we compare its stability and convergence to those of the standard 
SPH implementation using a set of idealised galaxy formation 
simulations of varying resolution. We will see that the new 
implementation retains a larger fraction of hot gas, avoiding 
the artificial overcooling often caused by inappropriate density estimates. 
The same test problem also demonstrates the numerical stability 
of the scheme in the presence of steep density (and temperature) 
gradients as well as in an accretion shock.

As noted above, proper modelling of the formation and evolution 
of a galaxy requires many physical processes to be considered in 
addition to the already complex interaction of nonlinear
gravitational evolution and dissipative gas dynamics.
The interstellar medium (ISM) where gas exists
in a wide range of density and temperature states, may be
considered as a multiphase system 
resulting from the interplay of processes such as gravity, hydrodynamics,
star formation, stellar photo-heating, shocking by supernovae and
stellar winds, cosmic ray and magnetic field dynamics, chemical 
enrichment and dust formation (Field 1965, Cox \& Smith 1974,
McKee \& Ostriker 1977, Ferrara \etal 1995, Efstathiou 2000).
Each process introduces its own
length and time scales which often differ by orders of magnitude
from those of the galaxy as a whole. As a result, a realistic
description of the galactic environment is a severe challenge both
for theoretical modelling and for numerical simulation.

Simulations of the interstellar medium which attempt to follow
all or many of these effects are only possible at the cost of 
studying a very small portion of a galaxy. The
formation of the galaxy as a whole cannot then be considered. A
small simulated region makes it possible to resolve
the few parsec length scale characteristic of the evolution
of individual expanding supernova remnants (see e.g. Rosen \& 
Bregman 1995, Wada \& Norman 2001, Avillez 2000). In order to 
achieve sufficient resolution, a two dimensional geometry is 
often adopted and important ingredients, for example magnetic
fields and cosmic rays, are often neglected. The results are 
thus still far from realistic. Our aim is different. We seek
to follow the overall evolution of a forming galaxy. We therefore
include small-scale processes only insofar as they
affect evolution on scales larger than a few hundred parsecs.
Such processes are included with an appropriate (and heuristic)
``sub-grid'' model. In particular, we focus our attention on
star formation and on feedback from supernovae and stellar winds,
adopting simple recipes similar to those commonly used in 
semi-analytic models for galaxy formation (White \& Frenk 1991, 
Kauffmann, White \& Guiderdoni 1993, Cole \etal 1994, Somerville 
\& Primack 1999).

In recent years, a number of authors have worked along similar
lines, coupling smoothed particle hydrodynamics codes with simple 
star formation and feedback prescriptions in order to study 
galaxy formation and evolution (Katz \& Gunn 1991, Katz 1992, 
Navarro \& White 1993, 1994, Steinmetz \& M\"uller 1994). Others have
implemented similar recipes in grid-based hydrodynamics codes 
for the same purpose (Cen \& Ostriker 1992, 1999).
This work has shown that, except at extremely low 
resolution, the implementation of feedback as a localised 
heat source is ineffective in regulating star formation (Katz 
1992). This is because most of the gas heated by supernovae is 
so dense that it radiates the injected energy immediately 
without significant effect on the dynamics. As a consequence, 
nothing prevents more distant and diffuse gas from cooling 
and adding itself to the rapidly star-forming, dense ISM. 
Such ``thermal feedback'' thus fails to produce the 
starburst-driven winds which can drive galactic fountains or blow gas out of weakly 
bound systems like dwarf galaxies. Although such winds
are observed and are thought to play an important 
role in galaxy formation, a robust numerical scheme 
which can generate them is still lacking. 

{\it Ad hoc} solutions of varying complexity have been proposed 
to remedy this weakness of the standard method (Navarro 
\& White 1993, Yepes \etal 1997, Hultman \&  Pharasyn 1999, 
Thacker \& Couchman 2000, Springel \& Hernquist 2002b).
Our own scheme is a further attempt in this direction, based
on an explicit separation of the protogalactic gas into
diffuse and dense (star-forming) components. Feedback energy
is used to heat the two components separately and to convert
gas from the dense to the diffuse component. As a result, it becomes
possible to cycle material between the two phases and
to heat the diffuse phase directly. Our scheme is thus able
to regulate star formation and to drive galactic fountains or winds.
In our view, it contains fewer arbitrary elements and is more robust than
earlier suggestions, and, in addition, it is closely related to the
modification of SPH which we propose in order to avoid unphysical
effects near phase boundaries. 

Our methods for simulating stellar feedback 
are described in \S 2.3. Their ability to suppress star formation, to
reheat cold gas and to drive outflows from the galactic disk 
is demonstrated in \S 3.3, where we also show how feedback
has a relatively larger impact in small mass galaxies.
Finally, in \S 3.4 we use our methods for a preliminary study of
the formation of galaxies in a $\Lambda$CDM universe. Despite their
relatively low numerical resolution, these simulations test our schemes in the
more realistic context of hierarchical
aggregation in an expanding universe. In this situation also we find 
that feedback suppresses star formation much more effectively than 
would be inferred from simulations employing more standard SPH methods.

\section{Description of numerical methods.}

Studies of galaxy formation and evolution require appropriate 
numerical techniques. These must follow not only the gravitational 
evolution of the system, but also both the dissipative 
hydrodynamics of systems with a very wide range of 
physical conditions, and the ``subgrid'' physics of star formation and feedback.
Our work is based on modifications of a software package named
GADGET. This is a tree-based N-body/SPH code written in
C and presented in Springel \etal 2001 (hereafter SYW). It is
publicly available in both serial and parallel versions, and we 
refer the interested reader to SYW for a general introduction. 
As we document in the following sections,
we have at several points made choices different from those
recommended as default by SYW. In particular, we have fixed 
$\alpha=3/2$ (rather than the recommended $\alpha=2$)
in the general form of the SPH momentum equation (Monaghan, 1992):
\begin{equation}
\frac{dv_{i}}{dt} = -\sum_{j=1}^{N} m_{j} \left(
                    \frac{P_{j}}{\rho_{i}^{2-\alpha}\rho_{j}^{\alpha}} +
                    \frac{P_{i}}{\rho_{j}^{2-\alpha}\rho_{i}^{\alpha}} +
                    \Pi_{ij} \right)\nabla_{i}W_{ij} ,
\label{sph_genmomeq}
\end{equation}
where $N$ is the number of neighbors, $P$ is the pressure, $\rho$ the 
density, $W_{ij}$ the symmetric smoothing kernel and $\Pi$ the 
artificial viscosity used to capture shocks (we define $\Pi_{ij}$ in 
more detail below). We found by experiment that $\alpha=3/2$ gives
better behaviour in the multiphase SPH scheme we present below.
Furthermore, we have changed the SPH energy equation to the 
form (for $\alpha=3/2$):
\begin{equation}
\frac{du_{i}}{dt} = \sum_{j=1}^{N} m_{j} \left(
		    \frac{(P_{i}/\rho_{i})}{\sqrt{\rho_{i}\rho_{j}}} +
		    \frac{1}{2}\Pi_{ij} \right)v_{ij}\cdot\nabla_{i}W_{ij} ,
\label{sph_genegyeq}
\end{equation}
where $u_i$ is the specific internal energy (basically the 
temperature) of particle $i$, and $v_{ij}$ is the relative velocity 
between particles $i$ and $j$. 
We found equation (\ref{sph_genegyeq}) to integrate
the thermal energy evolution with much less noise
than the manifestly symmetric form used in the public version of GADGET.

In the following subsections we describe other additions to the public 
version that are necessary to study problems involving radiative 
cooling, star formation and energy feedback from stars.

\subsection{Implementation of radiative cooling and multiphase SPH.}

As a first step towards a realistic simulation of 
galaxy formation,  we introduce radiative cooling processes into the
publicly available code. In our version of GADGET we adopt an
implicit solver for the cooling equation and we 
interpolate the cooling rates from tables published by Sutherland and Dopita
(1993) for a variety of gas metallicities, from
primordial ($H, He$ only) to supersolar ($[Fe/H]=+0.5$). We assume the
gas is always optically thin to its own cooling radiation.

When cooling is included, it represents an effective sink for 
the internal energy and pressure support of the gas.
As a consequence, steep density gradients develop, with 
the densest gas locked at the cutoff 
temperature for atomic cooling processes, i.e. at about 
$10^{4}$ K. Gas that remains at low density may be heated
to high temperature by accretion shocks, or may remain at the low 
temperatures  typical of unvirialised ``quasilinear'',
gas. Clearly, additional cooling and heating processes such 
as $H_{2}$ cooling (Abel \etal 1997), ionization by high energy 
cosmic backgrounds (Theuns \etal 1998) or stellar photo-heating 
(Wolfire \etal 1995) could lead to an even more complex, multiphase
structure with several characteristic density and temperature scales.

When such multiphase structure is present, standard implementations
of SPH overestimate the density for particles
that fall near the boundary of a higher density phase 
(Pearce \etal 1999). The usual assumption of small density gradients 
across the smoothing kernel breaks down in this regime, and nearby
clusters of high density particles cause an upward bias in the standard SPH estimate:
\begin{equation}
<\rho_{i}>=\sum_{j}^{N} m_{j} W_{ij}.
\label{std_dens_eq}
\end{equation}
Here $N$ is the number of neighbours $j$ of particle $i$  and $W_{ij}$ 
is a symmetric smoothing kernel. In order to avoid
this bias, which leads to artificially high cooling rates and
to spatial exclusion effects, we modify the neighbor search and the
density evaluation of equation (\ref{std_dens_eq}) in a way which
leaves the numerical scheme as simple as possible. Our scheme is
quite similar to that of Ritchie \& Thomas (2000, hereafter RT)
as far as the neighbour search is concerned, but is very different 
for all other aspects of the hydrodynamics. 

It is important to recognize that the local density is the gas property 
responsible for phase segregation (since it determines the local
cooling rate ). An appropriate density estimator for a particle of any 
given phase should use only local material which is also part of that 
same phase. Theoretical studies show that the coldest and densest
gas in a multiphase medium can contain a large fraction of the mass,
but has a very small filling factor (McKee \& Ostriker, 1977). Thus 
when evaluating SPH quantities for diffuse gas, it is important to
exclude dense phase particles which happen to lie within the
smoothing kernel. (The problem is less severe in the opposite direction
since few neighbours of a ``dense'' particle are typically part
of the diffuse phase.)  Thus gas at low densities should give dense 
neighbours a low ``importance'' (or weight in the terminology of RT)
to prevent them from monopolizing the sum in equation (\ref{std_dens_eq}).

In order to maintain the number of neighbours within a small $\Delta N$
around $N$, we assume that two particles $i$  and $j$ do not consider 
each other as neigbours if their densities are very different
(see condition (\ref{msph1_eq}) below). This would be
equivalent to RT's neighbour search method if the number of neighbours were
kept exactly constant. Allowing it to fluctuate within a small interval,
as happens in GADGET, makes essentially no difference. Notice that
unlike RT we do not make any assumption about pressure balance. Indeed, our
density estimate depends only on the positions and previous density
estimates of the particles. It is independent of their temperatures
and velocities. The importance of pressure gradients is evaluated
on a particle by particle basis, as we will explain below.

In detail, our new implementation of SPH multiphase gas works as follows.
The SPH particle $j$ is not considered as a neighbour of $i$, even if
their separation is small, when all the following conditions are fulfilled,

\begin{eqnarray}
& & \rho_{j}>A\rho_{i},         \label{msph1_eq} \\
& & u_{j}<u_{i}/A^{'},          \label{msph4_eq} \\
& & \mu_{ij} \simgt -c_{ij},  \label{msph5_eq} \\
\nonumber
\end{eqnarray}
and in this case $i$ is also deleted from consideration as a possible
neighbour for $j$.
Here $A$ and $A^{'}$ are large numbers (we assume in this work 
$A=A^{'}=10$), $u$ is the specific energy and other symbols are 
explained below. Condition (\ref{msph1_eq}) is our proposed solution 
to the density estimation problems outlined above; the additional
conditions are required to avoid occasional pathological behaviour.
In particular, a complication arises when two different phases, typically a cold, 
compact clump embedded in more diffuse gas, are not only in
contact but are also shocking together. In such situations a simple 
prescription based on a density criterion alone (or similar ones based on 
temperature) can produce severe integration instabilities as
the resolution is increased.

In SPH the conservation of 
energy and momentum requires symmetric pressure and 
artificial shock forces to ensure that Newton's third law is
satisfied.
If particle $j$ can see $i$, but particle $i$ cannot see $j$, then 
we need to make sure that the reaction of $i$'s force on $j$ is indeed
small compared to the total force on $i$.
This is generally the case near phase boundaries (where this asymmetry
can often occur) except in shock regions where pressure gradients
are very large or the velocity field has a large divergence.
Hydro forces in SPH are evaluated using local estimates
of $\nabla P$ and  of the artificial
viscosity pressure term that usually dominates the hydro force
in shocks. In order to understand whether particle $i$ can really
ignore the presence of $j$ we need to study in more detail the SPH force term.
The reduced pressure force, i.e. the reduced pressure gradient on $i$ 
due to $j$, can be written in general SPH form as a function of the free 
power index $\alpha$ in the following way (see equation \ref{sph_genmomeq}):
\begin{equation}
f_{ij} = \left( \frac{P_{j}}{\rho_{i}^{2-\alpha}\rho_{j}^{\alpha}}
	+\frac{P_{i}}{\rho_{i}^{\alpha}\rho_{j}^{2-\alpha}} \right),
\label{msph2_eq}
\end{equation}
where $P=(\gamma -1) \rho u$ is the pressure of a perfect gas and 
$\gamma$ is the adiabatic index.
The relative error in the hydro force on particle $i$ arising from 
the neglect of the contribution from  $j$ can be estimated as:
\begin{equation}
{\cal E}_{ij} = \frac{f_{ij}}{Nf_{ii}},
\label{msph2bis_eq}
\end{equation}
This quantity can be minimized with respect to $\alpha$ and 
$(u_{i},u_{j})$, using the condition
given in equation (\ref{msph1_eq}). One then finds that, in order 
to keep very small ${\cal E}_{ij}$, the best value for $\alpha$ is
$3/2$. We also require that the internal energy of $j$ should be 
small compared to that of $i$:
\begin{equation}
u_{j} << u_{i}.
\end{equation}
This leads to condition (\ref{msph4_eq}), which plays an important role
in collapse problems where the gas shocked via
accretion is more dense than the gas still in the infall phase.

Finally, we also need to consider the artificial viscosity term, $\Pi_{ij}$.
This dominates in shock regions and is defined as
\begin{equation}
\Pi_{ij}=-\alpha_{v}c_{ij}\mu_{ij}+2\alpha_{v}\mu_{ij}^2,
\label{visc1}
\end{equation}
where $\alpha_{v}\approx 1$, $c_{ij}$ is the average sound speed and
\begin{equation}
\mu_{ij}=\min \left [ h_{ij} \frac{ v_{ij}\cdot r_{ij} }
                       { r_{ij}\cdot r_{ij} + 0.01h_{ij}^{2} }
               ;~0 \right ].
\label{visc2}
\end{equation}
Note that $v_{ij}\cdot r_{ij}$ and thus $\mu_{ij}$ are always negative 
in shocks. (See SYW for more details on the implementation of
artificial viscosity in GADGET.) To avoid non-physical behaviour in
shocks we thus impose a ``non-shock'' condition, i.e. we require
that the particles pairs to be decoupled should not to be part of a 
shock front:
\begin{equation}
\Pi_{ij}<f_{ij}.
\label{msph5x_eq}
\end{equation}
This translates into condition (\ref{msph5_eq}). When the
conditions (\ref{msph1_eq}), (\ref{msph4_eq}) and (\ref{msph5_eq}) are all
satisfied, particle $i$ can safely 
ignore the much denser particle $j$ (and in order to obtain a more
symmetric formalism, we also ensure that $j$ 
ignores $i$ during its own neighbour search).

The scheme described in this section produces appropriate density 
estimates for particles in each phase of a multiphase system, and it
conserves energy and momentum just as well as standard SPH, even in 
extreme dynamical situations Note 
that the density jump across a nonradiative shock can be at most $4$ 
for $\gamma=5/3$ (here $\gamma$ is the adiabatic index). For such
shocks our scheme should produce exactly the same results as standard
SPH since condition (\ref{msph1_eq}) is never fulfilled. We have
checked that this is indeed the case (even at low resolution) in 
simulations of the self-similar cosmological infall model of 
Bertshinger (1985) as well as in several other nonradiative problems 
studied during the development of the numerical scheme.

\subsection{Star formation scheme.}

When gas reaches high density in galaxy formation simulations, it is
generally assumed that it will begin to fragment and turn into stars.
However, current computational capabilities do not allow direct 
simulation of this process within a forming galaxy.
Thus, one is forced to adopt heuristic laws to
describe star formation in the same way as in semi-analytic
models of galaxy formation (Kauffmann, White \& Guiderdoni 1993, 
Cole \etal 1994, Somerville \& Primack 1999). One must first define 
the conditions under which a gas particle is eligible for
star formation, then calculate the rate at which its gas is 
converted into stars (the star formation rate, hereafter SFR).
Since our recipes are ``standard", we just list them below, and
refer the reader to Navarro \& White (1993, hereafter NW) and Thacker and
Couchman (2000) for a more detailed discussion.

The conversion of gas mass into stellar mass is possible
if and only if all of the following conditions are satisfied:
\begin{eqnarray}
& {\rm (a)} & \nabla \cdot  {\bf v} < 0 , \nonumber \\
& {\rm (b)} & \rho_{gas} > \rho_{*} = 5.0\times 10^{-26}~{\rm g cm}^{-3} , \nonumber \\
& {\rm (c)} & \tau_{sound} > \tau_{dyn}  , \nonumber \\
& {\rm (d)} & T < T_{*} = 4.0\times 10^{4}~{\rm K}  , \nonumber \\
\label{sfr_cond}
\end{eqnarray}
where $\nabla \cdot  {\bf v}$ is the divergence of the local velocity
field, $\tau_{sound}=h/c_{s}$ is the sound crossing time, 
$\tau_{dyn}=1/\sqrt{4\pi G \rho}$ the free fall time,
$\rho$ indicates density and $T$ temperature.
When all conditions listed above are fulfilled by a generic SPH
particle, the star formation rate is calculated from the 
following equation: 
\begin{equation}
\dot{M}_{stars}=c_{*}\frac{M_{gas}}{\tau_{dyn}}.
\label{sfr_eq}
\end{equation}
We have found that conditions (\ref{sfr_cond}) and equation (\ref{sfr_eq}) 
with $c_{*}=0.1$ reproduce fairly well the ``laws'' which appear to
regulate the observed star formation properties of galaxies
(Kennicutt 1998):
\begin{equation}
\Sigma_{SFR} = (2.5 \pm 0.7)\times 10^{-4}
		\left( \frac{\Sigma_{gas}}{M_{\odot}{\rm~pc}^{-2}} \right) ^{(1.4 \pm 0.15)}
		\frac{M_{\odot}}{{\rm yr~kpc}^{2}}.
\label{kenny_eq}
\end{equation}
Stars are represented numerically by reducing the mass of the star-
forming SPH particle and creating a new collisionless ``star'' particle.
It is not feasible, however, to spawn an independent star particle
for every minor star formation event. The number of star particles 
would then grow prohibitively, and the mass of each would be too low 
compared to the initial resolution, leading to numerical problems
such as two-body heating in encounters with dark matter particles.
To avoid this, we treat the gas particles as hybrid gas/star particles in the
way suggested by Mihos \& Hernquist (1994), where a gas particle can have
a ``hidden'' stellar mass that contributes to its inertia but does 
not play a role in the hydro equations. Once this ``hidden'' stellar 
mass grows to some fixed fraction (here assumed $0.3$) of the 
initial particle mass, a new stellar particle is created and the mass 
of the original particle is reduced. 
Furthermore, when the mass of the SPH particle falls to $10\%$ of its
initial value, the leftover gas is shared  among gas neighbours.
If a particle is not itself receiving material from a dissolving 
neighbours, it will form three stellar particles and then dissolve. 
In practice, since star-forming gas particles are normally highly 
concentrated in small subvolumes of the simulation, this is true 
for most of them. Note that creating a stellar particle and dissolving 
a gas particle requires care in order to conserve mass, momentum and
energy, as well as other gas/star properties such as metallicity and 
stellar age, if these are being followed.

\subsection{Feedback from star formation.}

In this section we will explain our numerical scheme to distribute the energy
produced by star formation activity, the feedback. In principle, along
with feedback one could also study the production and ejection of
heavy elements. This would require the specification of stellar yields,
along with an initial mass function, in order to determine the
appropriate source terms from our dynamically determined star formation
rates. In the present study we will not, however, model the metal 
enrichment of the gas, but rather focus our attention on the energetics
of feedback. The amount of energy injected by each newly formed 
solar mass of stars is then the only quantity we need. We assume the same
value as NW:~$e_{sn}=4.0\times10^{48}{\rm~erg}~M_{\odot}^{-1}$. In
addition, we neglect the delay of a few tens of millions of years
between the formation of a stellar population and the injection of the
bulk of its feedback energy.

Feedback comes primarily from massive stars ($\simgt 8 M_{\odot}$) that
explode as supernovae and energise interstellar gas. Again, resolution
constraints do not allow this process to be studied in detail so it
must be modeled using simplified rules. The most straightforward way
to introduce supernova feedback is through injection of thermal energy
into neighboring gas particles in proportion to the SFR. This 
proves ineffective, however, since cooling rates are extremely high in 
the vicinity of all eligible sites for star formation; SPH gas
particles in these regions have, by definition, very high density (Katz 1992). 
In reality the ISM structure in star-forming regions is more complex, with
gas at a wide range of densities and temperatures. This well known 
``multiphase structure'' is a consequence of  a wide variety of 
interacting physical effects, many of which are not represented in our model. 
Our simulations do produce coexisting 
``cold dense'' and ``hot sparse'' phases, but these are a very crude
representation of the true ISM structure. Thus, we need a
method to include the energy input in our simulations which accounts
approximately for all the unresolved
physics that we cannot study directly (see Yepes \etal 1997,
Hultman \& Pharasyn 1999, Thacker \& Couchman 2000, Springel \& 
Hernquist 2002b for erlier approaches to implementing feedback to
a multiphase gas).

Here we propose a  new method for funneling feedback energy into the
ISM which takes explicit account of the two phase structure which
arises naturally in SPH simulations with cooling. We will loosely
refer to the two phases as ``cold'' and ``hot''. We define them as 
follows: the cold gas has $\rho > 0.1\rho_{*}$ and $T<2T_{*}$, 
while the rest of the gas will be considered ``hot'', even though in
cosmological simulations much of it will be cold, unshocked material
which has yet to fall into a dense system. When new stars are formed, 
we distribute the stellar feedback energy to neighboring gas particles 
with a fraction $\epsilon_{c}$ going to cold gas and a fraction 
$\epsilon_{h}$ to hot gas. Our fundamental hypothesis is that at large
scales (of ${\rm~kpc}$ order, say) the nett effect of all the 
complex {\it ``microscopic"} processes is well described by an energy input
shared by the {\it ``macroscopic"} phases in given proportions.
Values for $\epsilon_{c}$ and $\epsilon_{h}$ could be fixed from a 
complete theory of the ISM, describing all the relevant processes, or
through direct numerical simulations covering the wide range of scales
and physical conditions appropriate for star formation within galaxies.
Such numerical studies are not currently available and there is no
consensus on a specific analytic model. We therefore prefer to leave 
our feedback parameters as freely adjustable and to try to understand
how their values impact the problem we are interested in here, namely
the large-scale structure of forming galaxies.

From the numerical point of view, implementing our procedure requires an
additional neighbour search for each star-forming particle to identify 
separately its cold and hot neighbours. In GADGET, this is easily
accomplished through a modification of the neighbour search routines 
(see SYW for details). Feedback to
the hot phase is implemented by adding thermal energy to the ten
nearest hot neighbours. Feedback to the cold phase is instead 
accumulated in a reservoir within the star-forming particle itself, 
which is always a cold particle according to our definition and to 
the star-formation conditions given in equation (\ref{sfr_cond}). 
This continues until the accumulated energy is sufficient to heat 
the gas component of the particle above a certain threshold that we take 
as $50T_{*} \approx 10^{6}$ Kelvin. This is far enough above 
$T_{*}$ for the promoted particle to be considered ``hot'' in its 
subsequent evolution, and, more physically, is similar to the temperature
of the hot phase in the McKee \& Ostriker (1977) three-phase  model of
the ISM. If the mean temperature of the ten nearest hot neighbors
is higher than this value, we instead take twice their mean
temperature as the threshold for promotion. Again, this is
designed to ensure that a promoted particle stays ``hot'' for at least
as long as other nearby hot particles.

When this temperature threshold is reached, the energy from the particle
reservoir is dumped in its internal energy and the particle is ``promoted''
to be a hot particle. A new SPH density is then calculated
excluding all its cold neighbours. At the same time any ``hidden''
stellar content is dumped to these cold neighbours.
In this way, the density and the cooling rate are strongly reduced
for the promoted particle, its entropy is raised above that of 
the surrounding hot gas, and its phase is changed completely.

\section{Tests of the numerical techniques.}

In this section we carry out a number of tests in order to check
the stability and efficiency of our new schemes for multiphase gas
(hereafter MSPH) and for feedback (hereafter MFB). In most cases, 
the initial conditions for these tests are the rotating, centrally 
concentrated sphere described in NW which consists predominantly of 
dark matter ($90\%$ by mass). Superposed SPH and dark matter particles 
are placed on a grid and then perturbed radially to give a density 
profile of the form $\rho(r) \propto r^{-1}$. Velocities are chosen 
so that the sphere is initially in solid-body rotation with  
spin parameter $\lambda\equiv J|E|^{1/2}/GM^{5/2} \approx 0.1$.
The initial thermal energy of the system is always a negligible fraction
($\approx 5\%$) of the gravitational energy.
This system collapses from the inside out to form a disk galaxy as the
simulation proceeds. As an additional test in a more realistic
situation, we also carry out a low resolution study of the formation 
and evolution of a galaxy within in a $\Lambda$CDM cosmology.

As we stated above, we do not expect our MSPH scheme to give different 
results from standard SPH for non-radiative problems. We have checked
that this is indeed the case for these two kinds of initial
conditions. Thus, in the following we will only show tests in which 
gas is allowed to cool radiatively in order to study how MSPH and
MFB compare with standard SPH implementations in the presence of
multiphase structures.

\subsection{General numerical parameter settings.}

The code GADGET requires a number of numerical parameters to be
specified before a run can be executed. Particularly relevant here 
are parameters related to timestep and gravitational force accuracy. 
Softening lengths are quoted explicitly in the following subsections;
in general, we choose the softening length for the gas to be always half 
that for the dark matter, while the softening for the stars is the mean of the 
two. The timestep criterion we use is derived from the softening length 
as follows:
\begin{equation}
\delta t_{dyn}=\sqrt{2\eta\frac{\epsilon_{softening}}{a}},
\end{equation}
where $a$ is the acceleration and $\eta$ a constant that we fix to $0.08$.
For gas particles we also limit the timestep by a Courant criterion.
Our numerical implementation of the
Courant condition is that reported in SYW, where
we fix the Courant accuracy parameter to be $\alpha_{courant}=0.1$.
Finally we require timesteps to stay between $10^{-6}$ 
and $10^{-8}$ internal time units (our length, mass and velocity units 
are: Mpc/h, $M_{\odot}/h$ and km/sec). The lower bound is never
reached in any of the simulations shown here and the typical timestep is
a few $10^{-2}$ Myr.

In GADGET, gravitational forces are calculated by means of a tree
algorithm. We adopt the new opening criterion suggested in SYW, and we 
set the dimensionless force accuracy parameter to 
$\alpha_{force}=0.01$ (see SYW for a full detailed description
both of the criterion and of the parameter).

Finally, the artificial viscosity parameter that we will use is equal to 
$\alpha_{v}=1.3$, larger than the value commonly adopted in other studies.
We found this higher value to be necessary to prevent interpenetration
of particles both in SPH and in MSPH, when using the shear-corrected 
version of the artificial viscosity force (Balsara 1995, Navarro \& 
Steinmetz 1997).

\subsection{A convergence study of multiphase SPH.}

In our first set of numerical simulations we fix the physical
properties of the system and vary the resolution (i.e. the number 
of particles) in order to study the stability and convergence of 
our MSPH scheme. We allow the gas to cool radiatively
because we are particularly interested in the dynamics 
of multiphase gas. In order to isolate such effects, 
we do not allow the gas to be converted into stars. (This allows 
the gas density to reach very large values, producing an extreme 
density range.)
We present results for six different simulations of the rotating, 
collapsing sphere, three for SPH and three for MSPH. Along each set
the  resolution increases with $2000, 4000$ and $8000$ particles 
of each type and the dark matter softening set to $2, 1.5, 1 {\rm~kpc}/h$,
respectively. In all cases the total mass of the system was 
$10^{12}M_{\odot}/h$ and its initial radius was $100 {\rm~kpc}/h$.

\begin{figure}[t]
\includegraphics[width=1.1\textwidth, height=0.8\textheight]{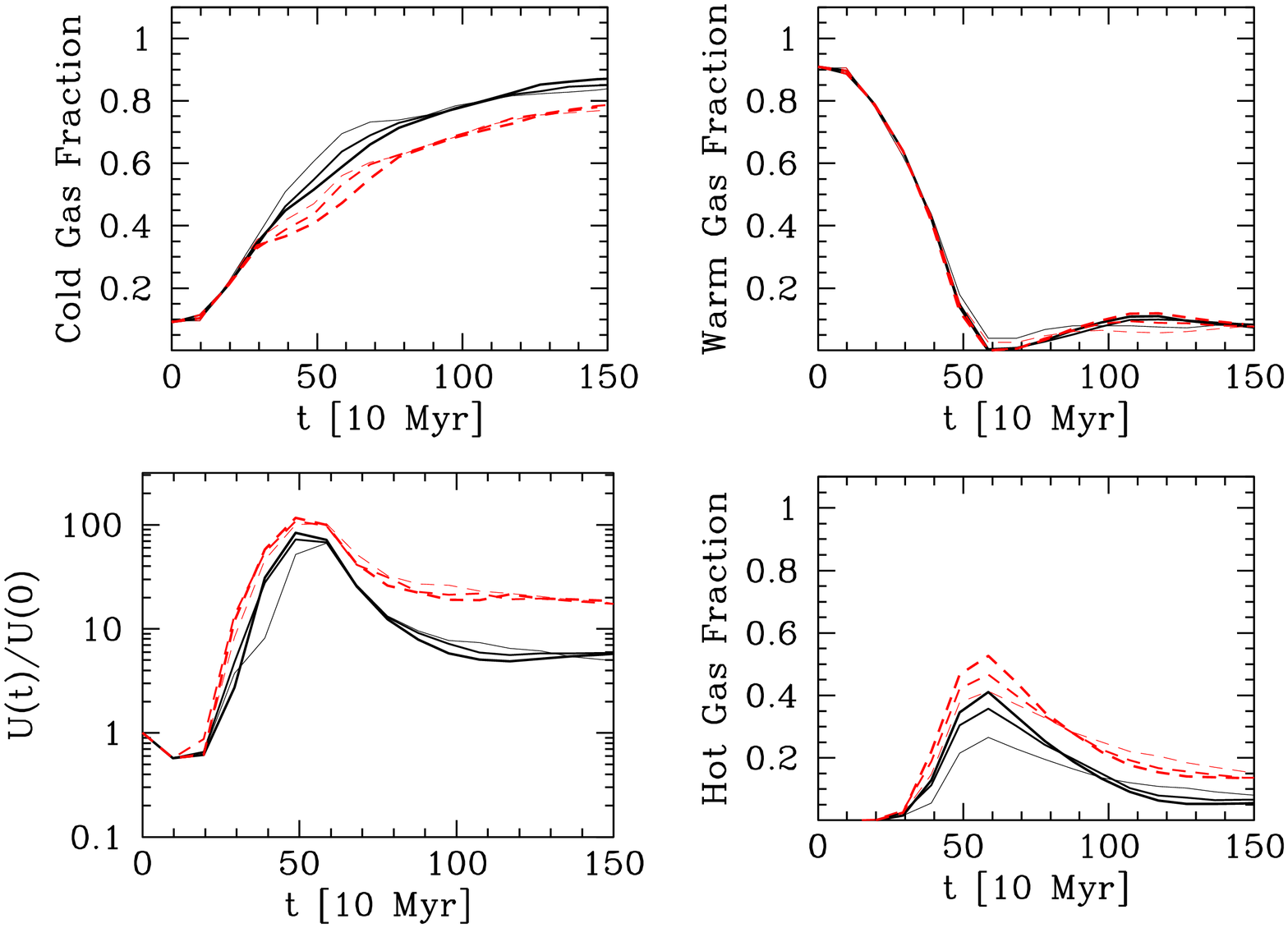}
\caption{\label{msph_ncnv_panel} Evolution of the total gas thermal
energy and of the gas mass fractions in various phases for
simulations
of the collapse of a centrally concentrated, rotating sphere made of
90\% dark matter and 10\% gas. Cooling is included in these
simulations but not star formation. The gas phases are defined as follows: 
{\it Hot:} $T\simgt10^{5}{\rm~K}$; {\it Cold:} $T\simlt10^{5}{\rm~K}$ and
$n_{H}\simgt0.1{\rm~cm}^{-3}$; {\it Warm:} otherwise. 
Solid lines give results for standard SPH, while dashed lines
are for our new MSPH scheme. The thickness
of the lines increase with the resolution of the simulation (2000,
4000, 8000 gas particles).
}
\end{figure}

In all these simulations an initial central collapse generates an 
accretion shock which moves out 
through the infalling envelope, while a fraction of the inner gas 
cools onto a dense and cold central core. At later times (after
about $30{\rm~Myr}$) this core becomes the centre of
a disk of cold, dense gas whose self-gravity
is balanced by rotation. The dark matter particles virialize violently 
under the effect of their self-gravity and finally assume a
centrally concentrated, spheroidal, quasi-equilibrium distribution.
In Fig. (\ref{msph_ncnv_panel}) we show the time evolution of the gas
fraction in each of three different phases, as well as the evolution 
of the total internal energy. Our definition of gas phases here
differs from that in Sec. 2.3 and is as follows:
\begin{itemize}
\item[]{\it Hot} when $T>2T_{*}$,
\item[]{\it Cold} when $T \leq 2T_{*}$ and $\rho>0.1\rho_{*}$,
\item[]{\it Warm} when $T \leq 2T_{*}$ and $\rho \leq 0.1\rho_{*}$.
\end{itemize}
Please notice that these definitions are unrelated to those of the
similarly named phases identified in thermal instability studies of 
the galactic ISM. In particular, our warm phase includes both
unshocked infalling gas and gas which has either shocked at low density or
has expanded adiabatically to low temperature after shocking.

As we can see in Fig. (\ref{msph_ncnv_panel}), cooling is less
efficient in the MSPH scheme than in standard SPH, since there is 
a clear increase of hot gas in MSPH runs and a decrease of the cold phase.
These differences are a result of excessive cooling in SPH at the interface between
the central cold disk and the surrounding hot atmosphere. This artifact
is caused by incorrect SPH density estimates for hot particles
near the disk and was first pointed out by Pearce \etal (1999). Note
that it is independent of the problem highlighted by Springel \& 
Hernquist (2002a) which arises from non-conservation of entropy in 
convergent flows for many SPH implementations. This latter problem 
is present to an equal extent in our SPH and MSPH models, and is less
severe than in the worst cases studied by Springel \& Hernquist (2002a)
because of our chosen representation for the SPH energy equation. 
Notice also that both schemes are quite stable in the regime investigated.

\begin{figure}[t]
\includegraphics[width=1.1\textwidth, height=0.8\textheight]{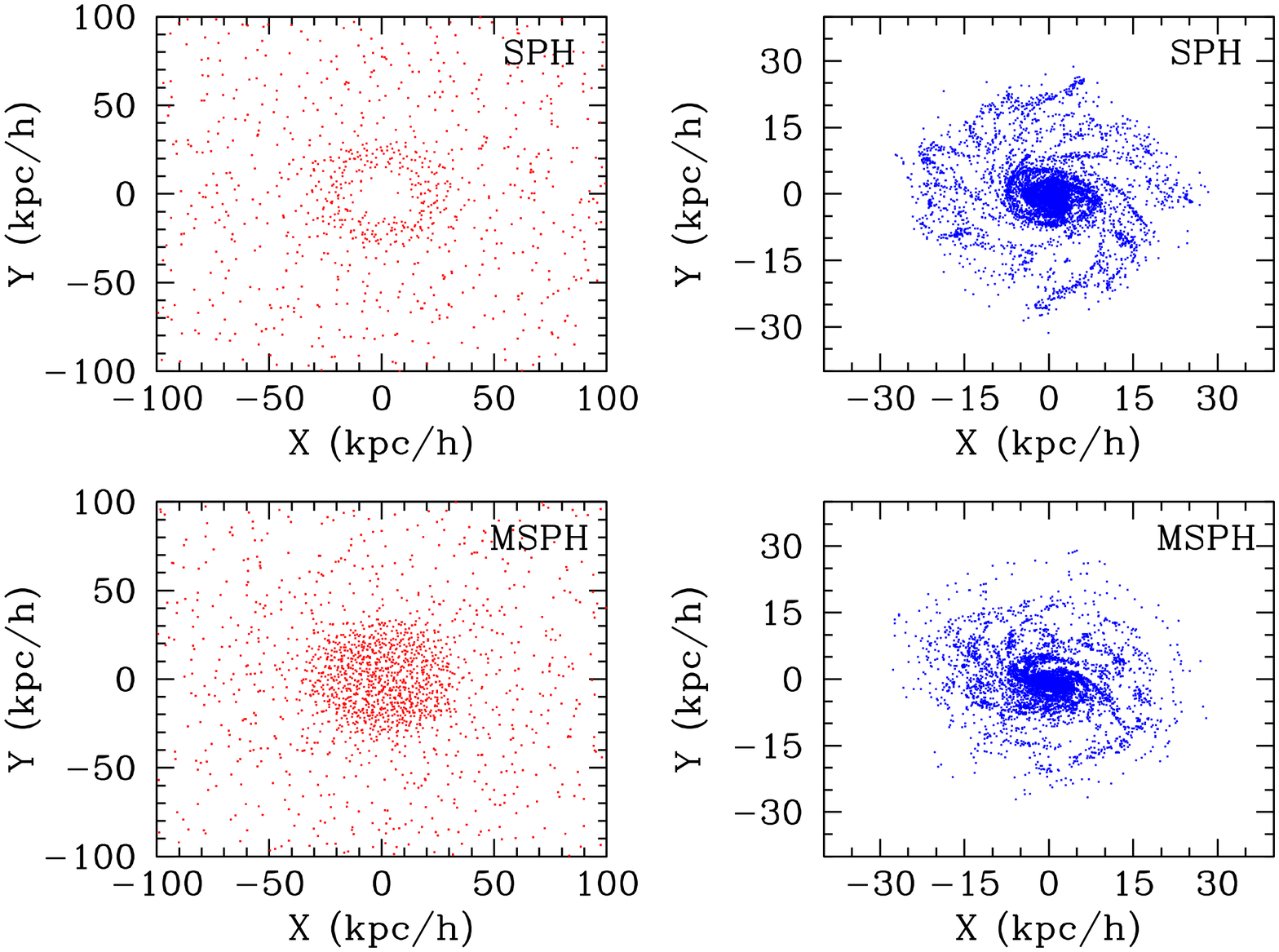}
\caption{\label{msph_ncnv_positions} Face-on projections at
$t\sim 1.2$ Gyr of our 8000 gas particle, cooling only simulations 
of a collapsing, rotating sphere. The standard SPH 
simulation is in the upper row with the MSPH simulation below it. 
The left-hand plots show ``hot'' particles ($T> 10^5$K) while the
right hand ones show ``cold'' particles ($T< 10^5$K and $n_H>0.1$
cm$^{-3}$).
}
\end{figure}

\begin{figure}[t]
\includegraphics[width=1.1\textwidth, height=0.8\textheight]{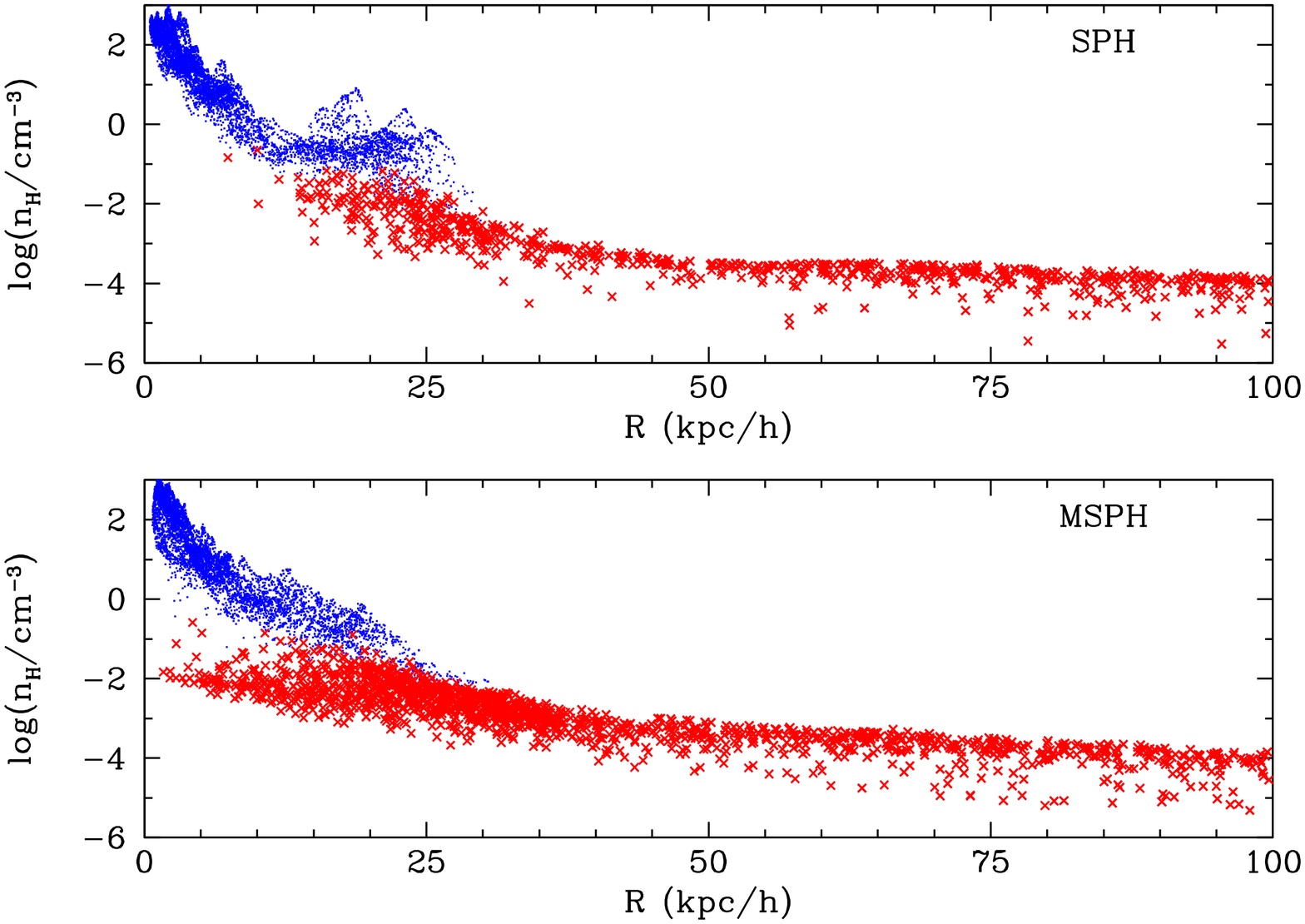}
\caption{\label{msph_ncnv_dprof} Particle densities as a function of 
galactocentric distance after $\sim 1.2$ Gyr of evolution in two 8000
gas particle, cooling only simulations of the collapse of a rotating 
sphere. The upper plot is for an SPH model and the lower for an MSPH
model. Gas particles with $T>10^5$K are plotted with crosses.
}
\end{figure}

A visual inspection of the particle distributions in Fig. 
(\ref{msph_ncnv_positions}) shows that hot particles do survive near the
central disk in the MSPH case. They have an almost spherically
symmetric distribution with density peaked at the centre. The cold
disk rotates within this ambient hot medium. In contrast, in the SPH
model hot particles are excluded from the vicinity of the disk so that
the hot phase actually has a density minimum at the centre of the
galaxy. This is seen most clearly in the gas density profiles of
 Fig. (\ref{msph_ncnv_dprof}). Such behaviour is clearly unphysical. 

\subsection{Star formation and feedback in an idealized galaxy formation problem.}

We now study the effects of our multiphase feedback (MFB) scheme, again
in the simple case of a rotating spherical collapse.

In the first set of simulations, we investigate how our scheme affects 
the evolution of total cold gas mass, total stellar mass and total
star formation rate in an object with $4000$ particles of each type, 
a dark matter softening length of $1.5 {\rm~kpc}/h$
and a total mass of $10^{12}M_{\odot}/h$. We present results for five
simulations. Four use MSPH with different values for the
adjustable feedback efficiency parameters $\epsilon_{c}$ and $\epsilon_{h}$, 
while the fifth is a control simulation using SPH and no feedback:
\newline
\newline
\begin{tabular}{cccc}
Name & $\epsilon_{c}$ & $\epsilon_{h}$ & Scheme \\
R01 & $0.0$ & $0.0$ & SPH  \\
R02 & $0.0$ & $0.0$ & MSPH \\
R03 & $0.0$ & $0.8$ & MSPH \\
R04 & $0.4$ & $0.4$ & MSPH \\
R05 & $0.8$ & $0.0$ & MSPH \\
\end{tabular}
\newline
\newline
The remaining fraction ($\epsilon=1-\epsilon_{c}-\epsilon_{h}$) of 
feedback energy is understood to be dumped in the classical way
as heat input to the cold gas. This energy is lost
by cooling and has no effect on the dynamics.

In a second set of simulations, we fix $\epsilon_{c}=\epsilon_{h}=0.40$ 
and study the effects of our MFB scheme on objects of different mass. 
We compare four runs with $4000$ particles of each type and with total 
masses $10^{9}M_{\odot}/h, 10^{10}M_{\odot}/h, 10^{11}M_{\odot}/h$
and  $10^{12}M_{\odot}/h$. We call these G09, G10, G11 and G12. Note
that G12 is identical to R04. The dark matter softening is $1.5 
{\rm~kpc}/h$ for the $10^{12}M_{\odot}/h$ object and is scaled down 
in proportion to $M^{\frac{1}{3}}$ for the others (as is the initial 
radius of the object, and so also its characteristic velocity in
virial equilibrium).

\begin{figure}[t]
\includegraphics[width=1.1\textwidth, height=0.8\textheight]{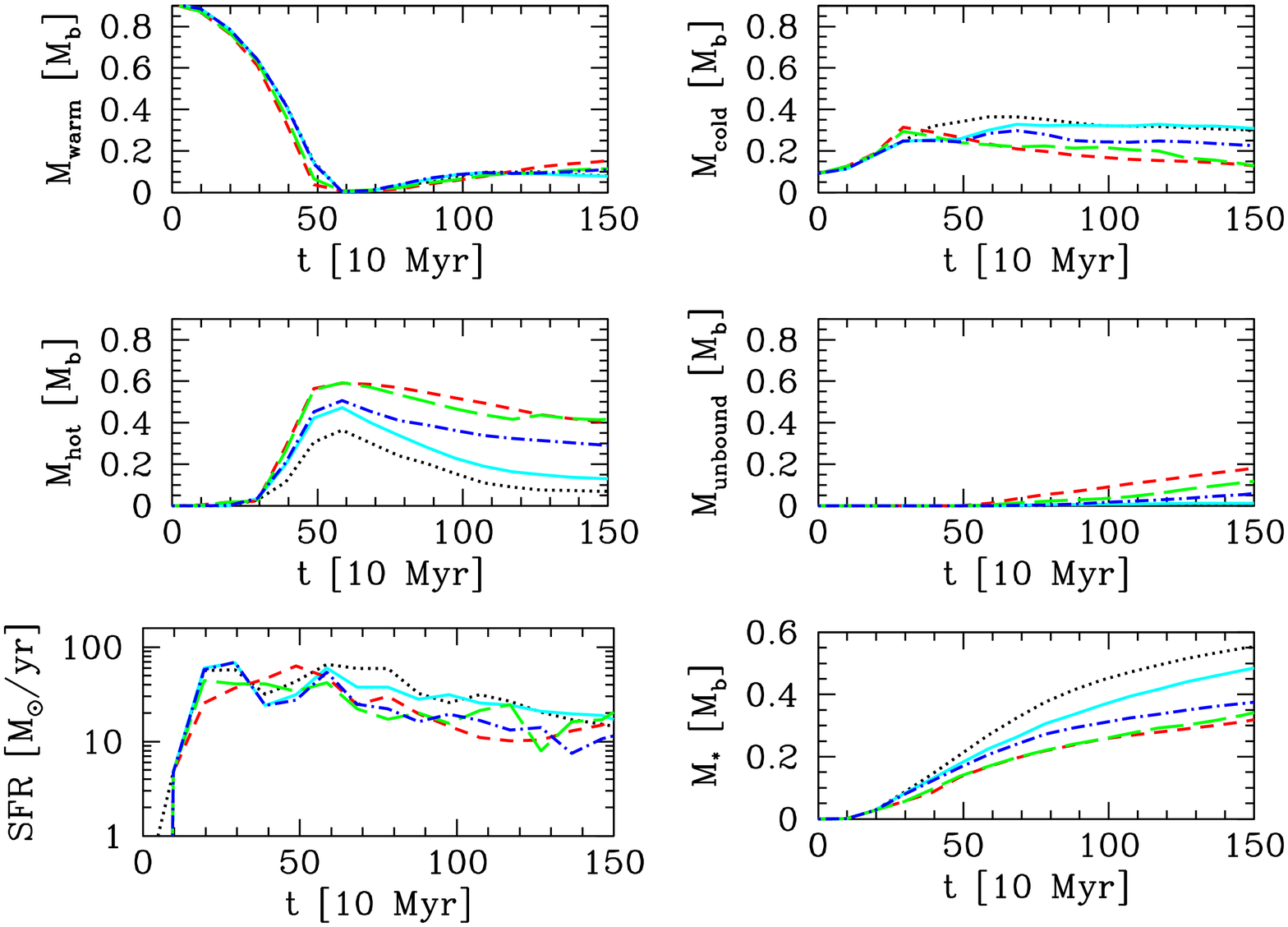}
\caption{\label{mfb_prfb_panel} The evolution of the star formation
rate and of the masses of various components in simulations R01
(dotted) R02 (solid) R03 (dashed) R04 (long-dashed) and R05
(dot-dashed). Their parameters are described in the text. Masses are 
given in units of $M_{b}$, the total baryonic mass
($10^{12}M_{\odot}/h$ for all these simulations). Hot, Cold and
Warm phases are defined as in figure (\ref{msph_ncnv_panel}), while 
the {\it unbound mass} is defined as the gas mass with positive
kinetic plus gravitational potential energy. The star formation rate, 
$SFR$, is given in solar masses per year, and $M_{*}$ is the total
mass in stars.
}
\end{figure}

\begin{figure}[t]
\includegraphics[width=1.1\textwidth, height=0.8\textheight]{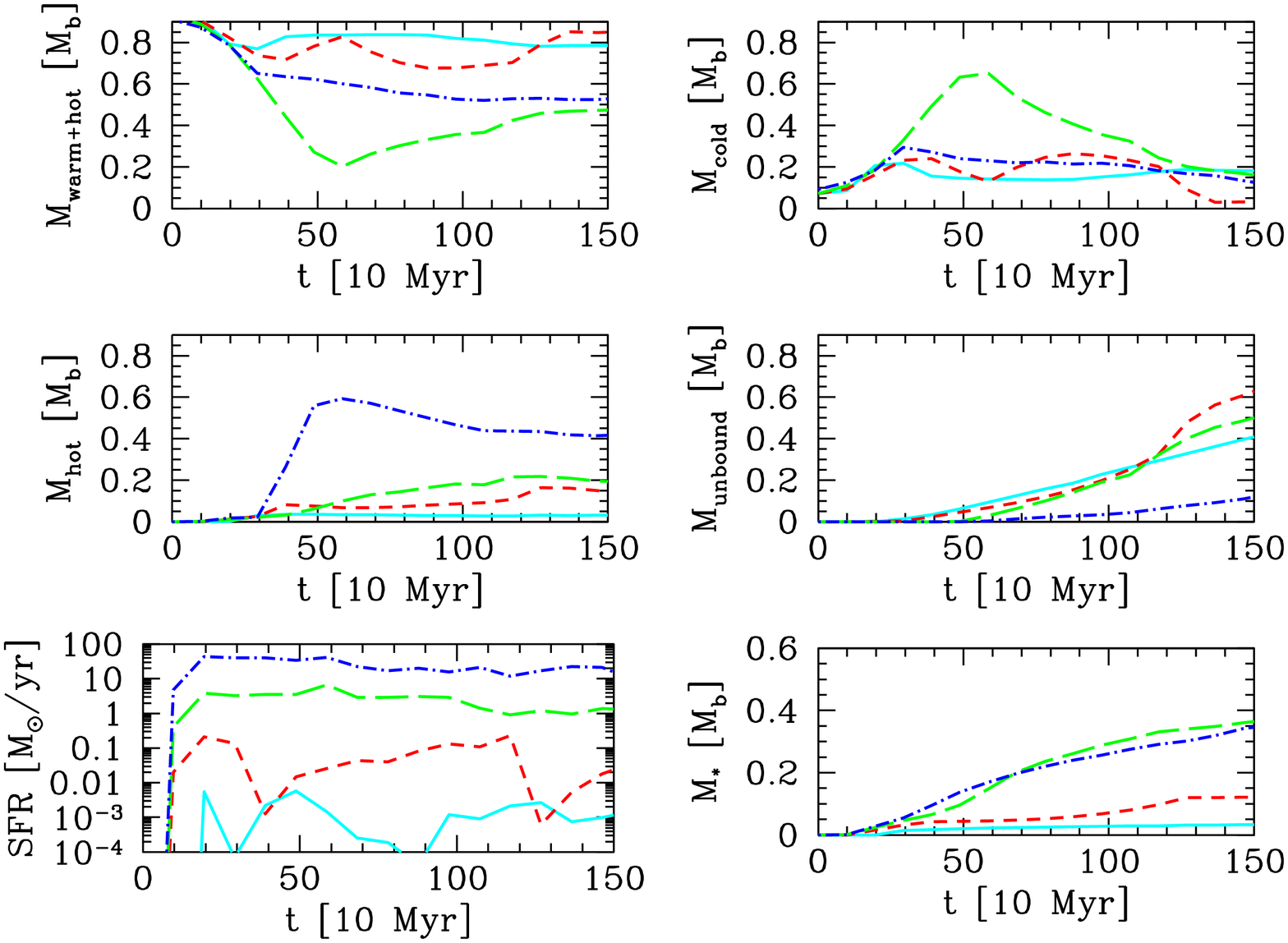}
\caption{\label{mfb_mass_panel} As Fig. (\ref{mfb_prfb_panel}) but for simulations  G09 (solid),
G10 (dashed), G11 (long dashed) and G12 (dotted-dashed). Again their
parameters are described in the text. For these simulations the total
baryonic mass $M_{b}$ is $10^{9, 10, 11, 12}M_{\odot}/h$ for G09,
G10, G11 and G12, respectively. Because of the varying virial
temperature of these systems we plot results only for the combined
warm+hot gas component.
}
\end{figure}

In Fig. (\ref{mfb_prfb_panel}) and (\ref{mfb_mass_panel}) 
we show the gas mass in various phases (definitions as before) 
in units of the total baryonic mass. The unbound fraction is defined
as the fraction of gas particles with positive  kinetic plus 
gravitational potential energy. Also shown is the evolution of the 
star formation rate and of the total stellar mass.

Runs R01 to R05 are shown in Fig. (\ref{mfb_prfb_panel}). The effect of 
varying the MFB parameters is most clearly seen in the evolution of
the hot, cold and  unbound fractions and in the total stellar mass.
A comparison of R01 and R02 shows how the better treatment of cooling
in MSPH already results in less star formation and a larger fraction
of hot gas. R03, R04 and R05 are all quite similar. 
Star formation is reduced by about 20\% from that in R02 in all three
cases. When $\epsilon_{h}$ is high it has the effect of driving
more gas out of the potential well, producing higher hot gas 
and unbound gas fractions. 

In Fig. (\ref{mfb_mass_panel}) we show results from models G09 to G12.
Note that since the characteristic virial temperature varies by a
factor of about 4 between neighboring models along this sequence, the
relative amounts of warm and hot gas by our previous definition are
not meaningful. For this figure we therefore plot 
the sum of the hot and warm fractions instead of the warm fraction alone.
The hot gas fraction evolves due to the initial accretion shock and 
the feedback energy input. Both cooling and feedback are relatively
more important in the low mass systems. The unbound gas mass is a 
good indicator of feedback effects. We see that less massive
objects allow a larger fraction of gas to be expelled from their
potential wells. This gas expands adiabatically and can cool to the
point where it is considered ``warm''. The specific star formation 
rate and the star mass fractions are also substantially reduced in
the low mass objects as a result of feedback. Thus conversion of
baryons to stars is about nine times less efficient overall in G09 than in
G12. Some care is needed when interpreting the evolution of
$M_{cold}$, since the amount of cold gas evolves not only through
cooling of the hot phase, but also through heating by feedback and
by conversion into stars. The last aspect is generally dominant.

The main limitation of these tests is the absence of substructure
or of ongoing infall onto the central galaxy. These simple initial 
conditions do not allow a proper description of outflows since the
expelled gas expands freely (and unrealistically) into the vacuum 
which surrounds the protogalaxy. In the next section we will study 
the more realistic case of a galaxy forming in a cold dark matter 
universe. Infall of clumpy material is then present at all times.

\subsection{Low resolution cosmological runs.}

As a final test we study the case of a Milky Way like galaxy forming
in a flat $\Lambda$CDM universe with $\Omega_{\Lambda}=0.7$, 
$\Omega_{b}=0.03$ and $h=0.6$. The initial conditions, kindly provided
by J. Navarro, consist of a high resolution region that contains the 
particles of the forming galaxy and its immediate environment.
This central region was excised from a large-scale simulation of a
cosmologically representative volume, the rest of which is represented
only through a distribution of relatively high mass dark matter
particles. These provide an appropriate tidal field on the high
resolution region (see Navarro \& Steinmetz (2000) for further
details). In the simulations we present here there are initially only 
7440 particles of each species in the high resolution region, about 
15\% of which typically end up in the dominant galactic halo. Our
adopted softening is $\epsilon_{gas}=0.5\epsilon_{cdm}=3.0
{\rm~kpc}/h$, and the initial mass of a gas particle is
$m_{gas}\approx 7.5 \times 10^{7}~M_{\odot}/h$. We use
exactly the same initial condition for three different simulations in
which the SPH scheme, star formation and feedback parameters are fixed 
as in the R01, R02 and R04 simulations described in Sec. 3.3; we will 
refer to these three simulations as SPH, MSPH and MSPH+MFB,
respectively. As a final experiment we repeat the MSPH+MFB simulation 
reducing the mass of the system by a factor of 100 and rescaling 
positions and velocities by a factor of $100^{-1/3}$ so that all
densities and freefall times are unchanged. The 
characteristic temperatures of objects then decrease by a factor of
$100^{2/3}$ so that cooling times are substantially shortened and 
feedback effects are much more pronounced. We refer to this model as
MFB/100. If MSPH+MFB corresponds to the formation of a galaxy of Milky
Way scale, then MFB/100 corresponds to the formation of a dwarf with
a characteristic rotation velocity of about 40 km/s.

Although the resolution of these simulations is quite poor, they allow
us to test our schemes in the presence of the expansion of the
universe and within the hierarchical evolution characteristic of 
cold dark matter models. Clearly, we cannot address in detail the 
structural properties of our ``galaxies'', but we are principally 
interested here in their global evolution, in particular, in 
their star formation history and in the distribution of baryons 
over the various phases. We note that although our principal galaxy 
is made from only $\sim 1000$ gas particles, this is still quite a large 
number in comparison with the number of gas elements which form into
a typical ``galaxy'' in most simulations of galaxy formation within 
a ``cosmologically representative'' volume.

\begin{figure}[t]
\includegraphics[width=1.1\textwidth, height=0.8\textheight]{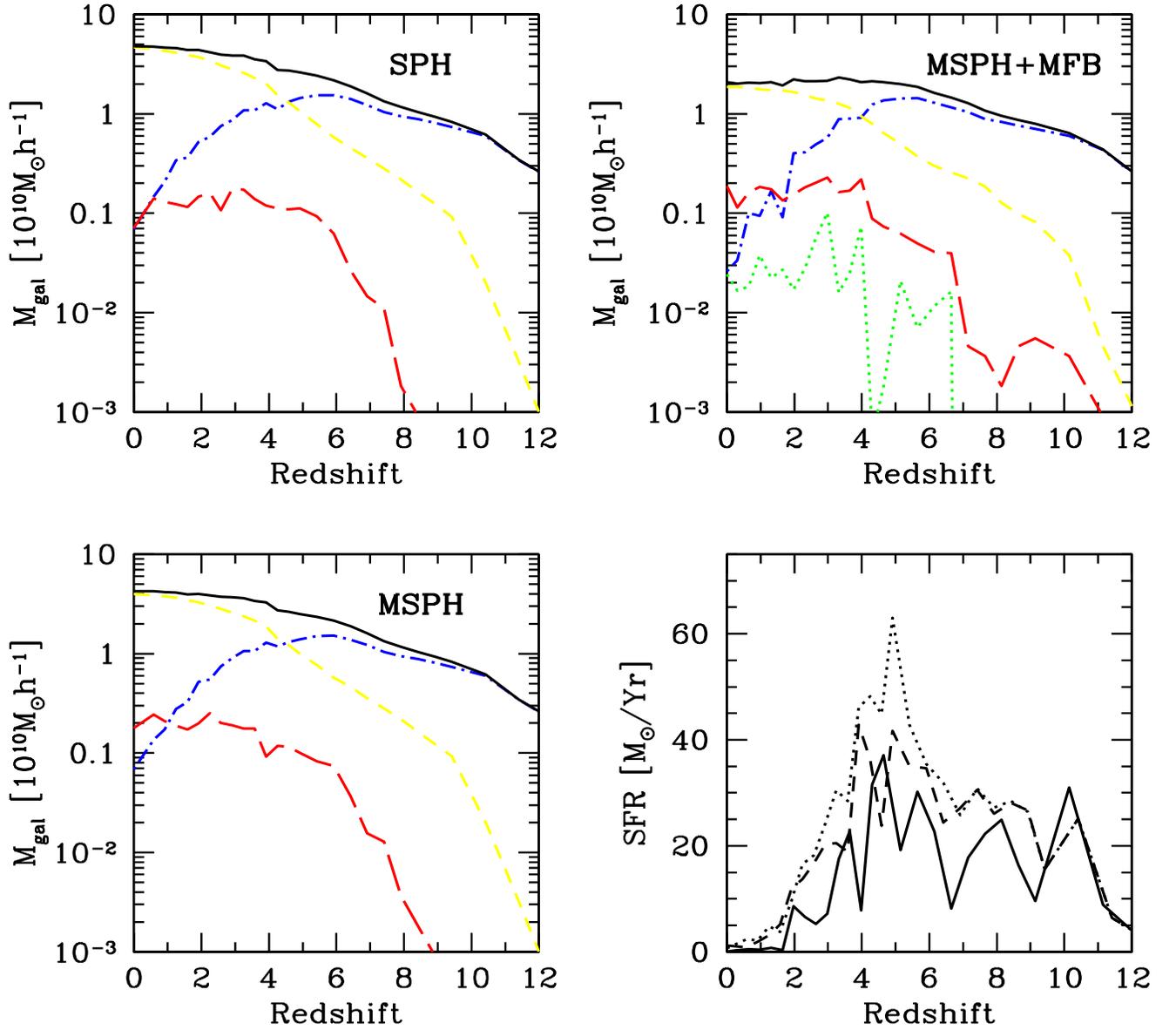}
\caption{\label{mfb_lcdm_panel} Evolution of the most massive galaxy in
three of our $\Lambda$CDM runs. The SPH model is at top left, MSPH at
bottom left, and MSPH+MFB at top right. In these panels
solid lines refer to total baryonic mass, dashed lines to stellar
mass, dot-dashed lines to the mass in cold gas, and long-dashed lines
to the hot+warm gas. For the MSPH+MFB run, the dotted line shows
the mass of currently diffuse gas which has been promoted from the cold phase
by feedback. The bottom right panel compares the star formation
histories of the three objects (dotted, dashed
and solid lines refer to SPH, MSPH and MSPH+MFB, respectively).
}
\end{figure}

In Fig. (\ref{mfb_lcdm_panel}) we display the evolution of various
components of the most massive galaxy in each of our three simulations.
We identify galaxies by applying a group finding algorithm named HOP
(Eisenstein and Hut, 1998) ti the baryonic components of the simulation.
We adopt the numerical parameters suggested by its authors.
HOP groups are characterized by 
boundaries that exceed a given overdensity threshold (we use $80$ times
the mean cosmic baryonic density).
However, the internal structure is also taken into account 
and local maxima (defined by a density peak threshold) which are connected by
thin bridges are split, according to the topology
of a higher density contour.
Eisenstein and Hut (1998) demonstrate that the HOP algorithm
is sensitive only to the density contrast on the boundary contour. We refer to their work
for a complete discussion of the methods and a comparison with other
popular group finding algorithms.

The plots show the evolution 
of the total baryonic mass, the stellar mass, the cold (dense) gas, 
the diffuse (warm plus hot) gas, and the successfully promoted mass 
(the diffuse mass at each time that was promoted from cold to hot at 
higher redshift and is still in the diffuse phase). Also shown is 
the star formation history of the most massive galaxy in each of
the three simulations. Note that because of HOP's overdensity criterion
only hot gas relatively close to the main galaxy is included in these plot.

A comparison of the SPH and MSPH models shows relatively modest
differences. As in the simplified models R01 and R02 of \S 3.2,
this is a consequence of the high efficiency of cooling and star
formation in the two schemes; more than 95\% of the baryons within the
main halo are converted into stars in both cases. The overcooling
problem with standard SPH is evident only at late times ($z<1$) 
when it results in a collapse of the hot atmosphere surrounding
the central galaxy and the accretion of about 10\% additional
gas onto the halo which rapidly cools and turns into stars. 

The inclusion of feedback causes much larger effects. Until $z\sim 5$ 
the total baryon content and the cold gas content of the MSPH and 
MSPH+MFB halos are almost identical, while star formation is slightly
reduced in MSPH+MFB. After this time, however, feedback has a
substantial effect on the total baryon content which remains
approximately constant in MSPH+MFB over a period when it doubles
in MSPH. Clearly the addition of baryons through accretion at $z<5$
is almost exactly balanced by outflow in a feedback-driven wind.
Thus by $z=0$ the stellar mass of the MSPH+MFB galaxy is a factor of
two smaller than that of the MSPH galaxy, and an equal mass of baryons
has either been driven out of the halo or prevented from accreting in
the first place. The star formation rate plot shows clearly that
the major effect of the feedback is to reduce star formation at late
times, not at the redshifts where the SFR is highest. This late
time star formation nevertheless accounts for most of the stars
in both models.

Further points to note from  Fig. (\ref{mfb_lcdm_panel})
are that the galaxies in all models are cold gas dominated at
redshifts above about 4 and are star-dominated at lower redshifts.
In addition, in the model with feedback only a small fraction of the
hot gas in the halo at any given time is made up of material that has
been reheated and evaporated from the disk by feedback. Most of the 
hot gas has yet to cool onto the disk for the first time.

\begin{figure}[t]
\includegraphics[width=1.1\textwidth, height=0.8\textheight]{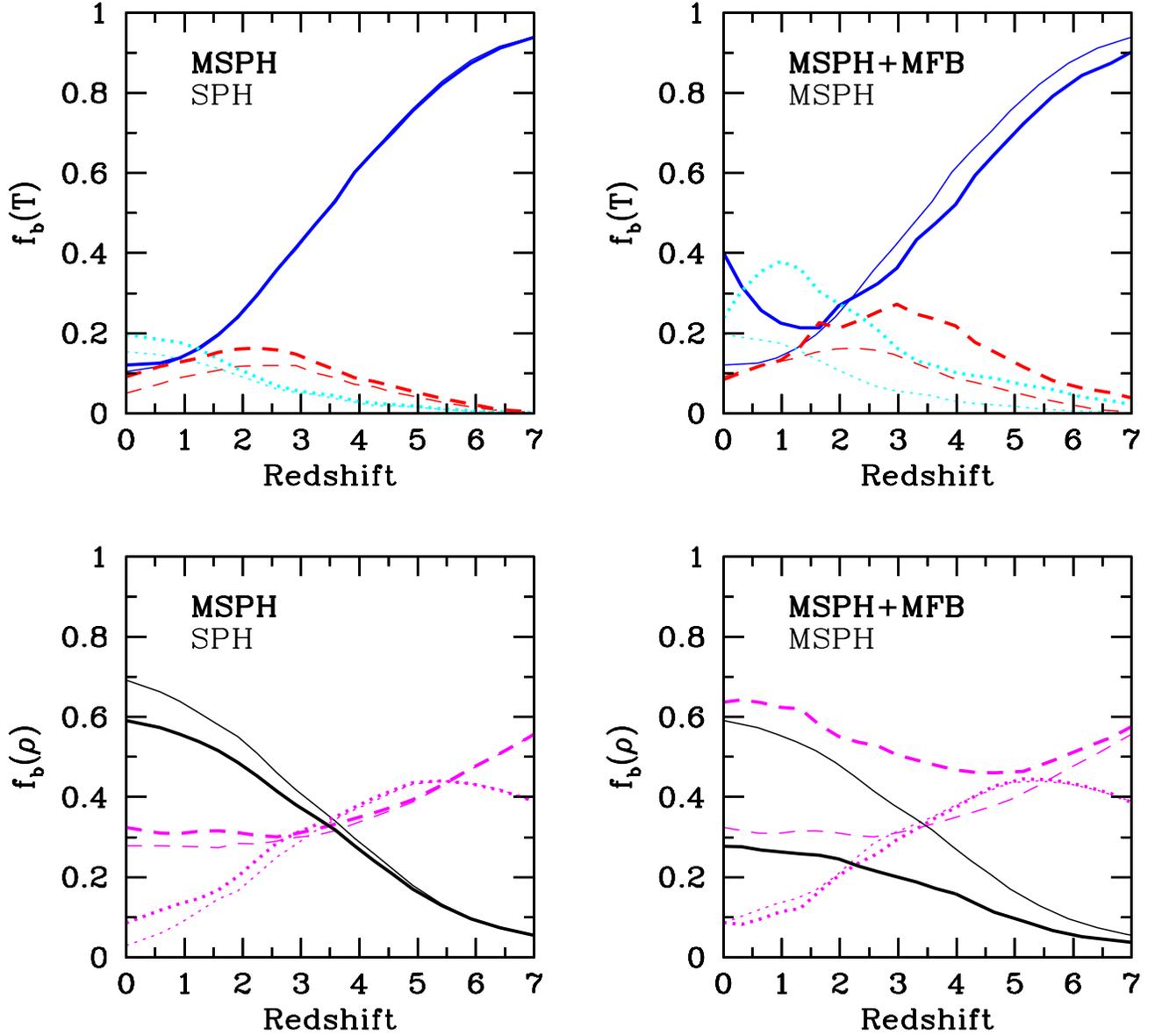}
\caption{\label{mfb_lcdm_fracs} Evolution with redshift of the baryonic
mass fraction in various components averaged over the full high
resolution region of our $\Lambda$CDM simulations. The left hand
panels compare MSPH (thick lines) with SPH (thin lines). The right
hand panels compare MSPH+MFB (thick lines) with MSPH (thin lines).
The upper plots show the mass fractions with $T<4\times
10^4$K (solid lines), with $4\times 10^4$K$<T<4\times 10^5$K (dotted
lines) and with $T>4\times10^5$K (dashed lines). The lower plots
show the stellar mass fraction (solid lines), the gas mass fraction at
densities more than 10 times the cosmic mean (dotted lines) and the
gas mass fraction at densities below this same threshhold (dashed lines).
}
\end{figure}

It is interesting to look for the missing baryons in the MSPH+MFB 
simulated galaxy  and to see where they end up. In Fig. 
(\ref{mfb_lcdm_fracs}) we show the evolution of the baryonic 
mass fraction in various temperature and density ranges, averaged over
the full high resolution region of our simulations. The left panels in 
this figure compare results for SPH and MSPH, the two models without
feedback. The evolution of the cold gas fraction ($T < T_*$) is nearly
identical in the two cases because this is almost all gas which has
yet to fall into any halo. The MSPH model has more warm and hot gas
and fewer stars at all redshifts (but most noticeably at low redshift) 
again reflecting overcooling near phase boundaries in the standard 
algorithm. The amounts of hot gas differ by more than a factor of two 
at $z=0$.

The right panels of Fig. (\ref{mfb_lcdm_fracs}) make a similar
comparison between the MSPH and MSPH+MFB models. Here the differences
are very large. Feedback reduces the overall amount of star formation
by about a factor of two. This reduction is compensated by a large 
increase in the amount of gas at low density; the amount of gas at densities greater than 10 times
the mean is almost the same in the two models. At redshifts above 
3 most of the excess diffuse gas in the MSPH+MFB model is at temperatures
above $10T_*$, at $0.5 < z < 3$ the bulk is at temperatures between
$T_*$ and $10T_*$, while at $z=0$ most of the excess is at
temperatures below $T_*$. Clearly, feedback driven winds are driving
gas into low density regions where it cools adiabatically as the 
Universe expands
 
\begin{figure}[t]
\includegraphics[width=1.1\textwidth, height=0.8\textheight]{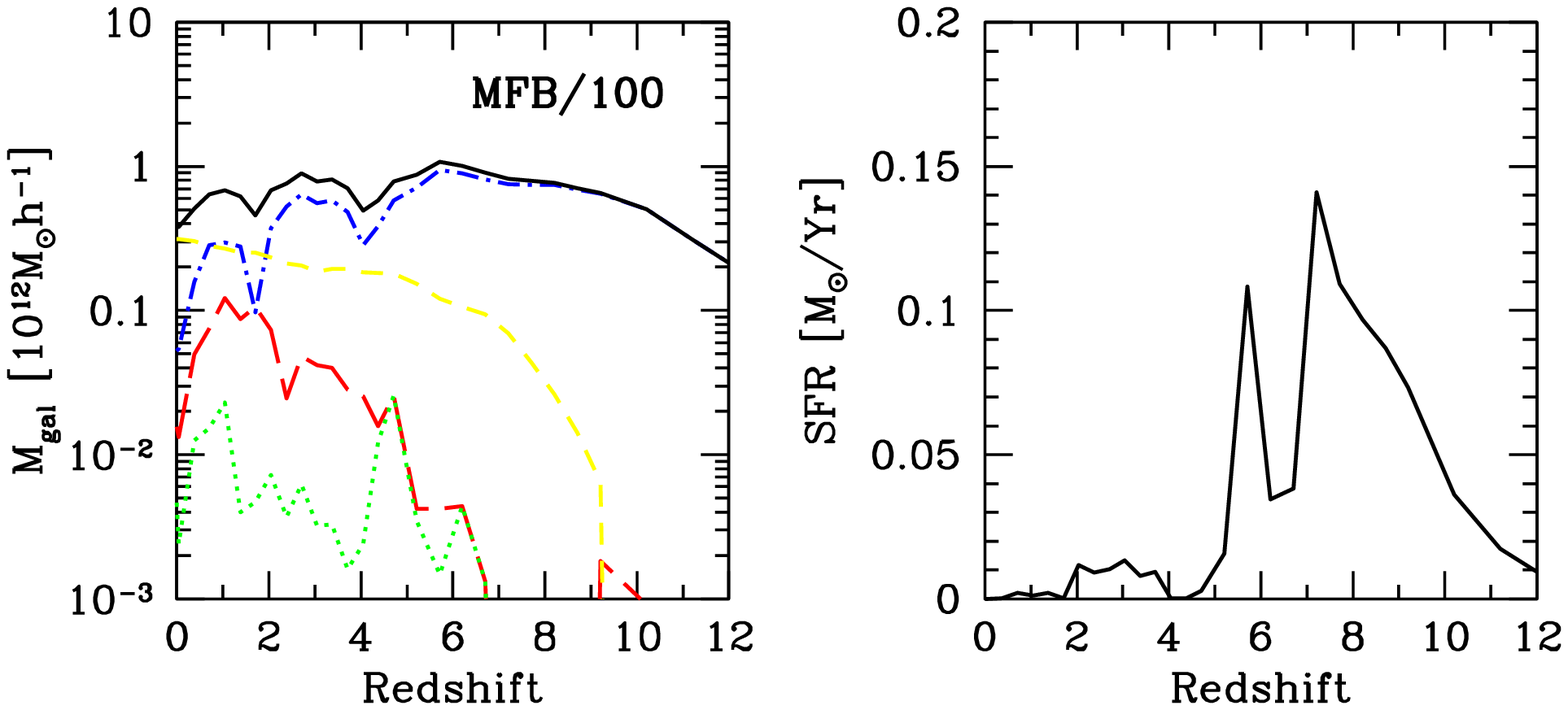}
\caption{\label{mfb_lcdm_resca} Evolution of the most massive galaxy in
the MFB/100 run. The lines and plots correspond to those of 
Fig. (\ref{mfb_lcdm_panel}) except that the mass unit is 100 times
smaller to account for the rescaling of the simulation.
}
\end{figure}

In Fig. (\ref{mfb_lcdm_resca}) we show plots similar to the right-hand
panels of Fig. (6) but for the MFB/100 simulation. In the absence of
feedback this model would behave almost identically to MSPH, since
cooling is efficient in most collapsed objects for both scalings of the
initial conditions. When feedback is included, its effects are much
more dramatic in MFB/100 than in MSPH+MFB because specific
binding energies are reduced by a factor of more than 20 while the
feedback energy injected per unit mass of new stars is unchanged. 
With these parameters star formation is much less efficient in the dwarf than in
the big galaxy. The dwarf is dominated by cold gas until redshifts
well below unity and it still contains more than 15\% cold gas at $z=0$.
For comparison the giant contains less than 2\% cold gas at this
time and is star-dominated for all $z<4$. The overall fraction of
baryons converted to stars is also greatly reduced in MFB/100. While 
about 50\% of the total available baryons get turned into stars in the 
giant, this is reduced to about 7.5\% in the dwarf. All these trends
agree qualitatively with observation, in that the gas fractions of
observed dwarfs are indeed substantially higher than those of giants,
and their much lower metallicities argue for a lower overall
efficiency of star formation and for ejection of large amounts of
gas by winds (e.g. Larson 1974; Dekel \& Silk 1986).

A further interesting and suggestive aspect of the MFB/100 simulation 
is the fact that its baryon mass oscillates and indeed decreases
overall from redshift 6 until the present. This is a result of the
interplay between accretion events, bursts of star formation, and 
the associated injections of feedback energy. Small galaxies do indeed
appear to have ``burstier'' star formation histories than giants
(e.g. Kauffmann \etal 2002 and references therein) so this behaviour
of the models may also be an echo of reality (see also Gnedin 1998).

To conclude, our MSPH scheme successfully reduces overcooling in
poorly resolved multiphase systems, and our MFB scheme allows feedback
to be effective in driving winds and in reheating cold interstellar
material, both processes which are observed to be important in
starburst systems. In addition, our MSPH scheme, in contrast to
traditional SPH implementations, results in a physically plausible 
spatial distribution for hot diffuse gas in the vicinity of a cold
dense ISM, while our feedback scheme, when applied to systems
of widely varying mass, leads to star formation efficiencies which
scale roughly as inferred from the observed metallicities and gas 
fractions of galaxies.

\section{Conclusion.}

In this paper we have proposed and tested two modifications of the
standard algorithms used for SPH simulations of the formation of
galaxies.  The first (MSPH) is designed to reduce artifacts which
occur in the common (and often poorly resolved) configuration of cold, 
dense gas clouds embedded in a hot diffuse halo. The second (MFB) is
a new implementation of feedback which allows supernova energy to be
channeled effectively into the heating of diffuse gas and the
evaporation of cold clouds.

When strong density jumps are absent, for example in most
non-radiative problems, our MSPH scheme reduces to a standard SPH
algorithm. In the presence of cooling a multiphase structure can arise,
and our scheme then eliminates the artificial overcooling discussed by
Pearce \etal (1999); particles in the hot phase which happen to lie
near a clump of cold gas have their density, and thus their radiative
cooling rate, substantially overestimated by the standard SPH formula.
In our scheme such cold, dense neighbours are not considered when calculating
densities for particles in the diffuse phase. This modification also
allows diffuse gas to take up a realistic spatial structure in
the presence of an embedded cold component. This is not the case for
standard algorithms (see Figs. (2) and (3)). Finally in strongly
dynamic situations our MSPH scheme conserves energy and momentum to
the same accuracy and is just as stable as standard algorithms when 
similar timestep criteria are used.

Since the work of Katz (1992) it has been recognized that implementations 
of feedback which simply inject supernova energy into the thermal
reservoir of neighboring gas particles have little effect on the
dynamics of SPH simulations; most of the energy is radiated
before it can accelerate the gas. Many alternative 
schemes have been proposed (e.g. Navarro \& White 1993, Yepes 
\etal 1997, Hultman \& Pharasyn 1999,
Thacker \& Couchman 2000, Springel 2000, Springel \& Hernquist 2002b) 
but none is yet accepted as a proper representation of the unresolved
``microphysics''. Our MFB scheme is original in several respects and 
is designed to facilitate reproducing the observed properties of
starbursts, while introducing as few {\it ad hoc} elements as 
possible. We use the supernova energy with predefined efficiencies to
heat the hot diffuse phase and to evaporate gas from cold clouds into
the diffuse phase. As our tests show, this not only allows feedback to
regulate star formation, but also generates winds or galactic fountains
without dialling in their characteristics ``by hand'' and permits such 
flows to entrain significant amounts of cold interstellar material. 

Our tests have concentrated on the idealised rotating, collapsing sphere 
of Navarro \& White (1993) and on the formation of a single isolated
galaxy and its environment in a $\Lambda$CDM universe. We have used 
relatively small numbers of particles in these experiments both to 
facilitate testing and because these kinds of algorithms are often
used to study galaxy formation within cosmologically ``representative''
regions (e.g. Katz. Weinberg \& Hernquist 1996, Pearce \etal 1999;
Murali \etal 2002, Springel \& Hernquist 2002a,b); a large fraction of the
``galaxies'' then form from fewer than (say) 1000 gas particles. Our
tests show that our proposed algorithms are numerically stable, and 
that for plausible choices of the heating efficiencies they  
reproduce the main qualitative features of observed
star-forming galaxies; self-regulation of star-formation; bursting
behaviour in small systems; the generation of fountains and winds
with simultaneous inflow and outflow; the entrainment of disk gas by
winds. The obvious next steps are to carry out much larger simulations both of
the formation of individual galaxies and of representative
volumes. The first can study the origin of the spatial, kinematic
and chemical structure of galaxies, checking whether
more realistic feedback can indeed solve the disk angular momentum 
problem (Navarro \& White 1994, Navarro \& Steinmetz 1997, Weil, Eke 
\& Efstathiou 1998; Thacker \& Couchman 2001). The second can study
how galactic winds enrich and structure the intergalactic medium.
We are currently pursuing projects in both these directions.

\newpage

\newpage

\newpage

\newpage

\newpage

\newpage

\newpage

\newpage


\newpage
\begin{references}

Abel T., Anninos P., Zhang Y., Norman M. L. 1997, NewA, 2, 181

Avillez M. A. 2000, MNRAS, 315, 479.

Balsara D. S. 1995, J. Chem. Phys., 121, 357 

Bertschinger E. 1985, ApJS, 58, 39

Cen R. \& Ostriker J. P. 1992, ApJL, 399, 113

Cen R. \& Ostriker J. P. 1999, ApJL, 519, 109

Cole S., Aragon-Salamanca A., Frenk C. S., Navarro J. F. \& Zepf S. E. 1994, MNRAS, 271, 781

Cox D.~P.~\& Smith B.~W.\ 1974, ApJL, 189, 105

Dekel A. \& Silk J. 1986, ApJ, 303, 39

Efstathiou G. 2000, MNRAS, 317, 697

Eisenstein D. J. \& Hut P. 1998, ApJ ,498, 137

Ferrara A., McKee C. F., Heiles C. \& Shapiro P. R. 1995, ASP Conf. Ser. Vol 80

Field G. B. 1965, ApJ, 142, 531.

Gingold R. A. \& Monaghan J. J. 1977, MNRAS, 181, 375 

Gnedin N. Y. 1998, MNRAS, 294, 407.

Hultman J., Pharasyn A. 1999, A\&A, 347, 769

Katz N. \& Gunn J. E. 1991, ApJ, 377, 365

Katz N. 1992, ApJ, 391, 502

Katz N., Weinberg D.~H. \& Hernquist L. 1996, ApJS, 105, 19

Kauffmann G., White S. D. M., Guiderdoni B. 1993, MNRAS, 264, 201

Kauffmann G., Heckman T.~M., White S.~D.~M., Charlot S., Tremonti C., Peng E.~W., Seibert M., Brinkmann J., Nichol R.~C., SubbaRao M. 2002, astro-ph/0205070, submitted to MNRAS

Kennicutt R. C. 1998, ApJ, 498, 541

Yepes G., Kates R., Khokhlov A. \& Klypin A. 1997, MNRAS, 284, 235

Larson R. B. 1974, MNRAS, 169, 229

Lucy L. B. 1977, ApJ, 82, 1013

McKee C. F. \& Ostriker J. P. 1977, ApJ, 218, 148

Mihos J. C. \& Hernquist L. 1994, ApJ, 437, 611

Monaghan J. J. 1992, ARAA, 30, 543

Murali C., Katz N., Hernquist L., Weinberg D.~H. \& Dav{\' e} R. 2002, ApJ, 571, 1

Navarro J. F. \& Steinmetz M. 1997, ApJ, 478, 13

Navarro J. F. \& White S. D. M. 1993, MNRAS, 265, 271 (NW)

Navarro J.~F.~\& White S.~D.~M. 1994, MNRAS, 267, 401

Pearce F. R., Jenkins A., Frenk C.S., Colberg J. M., Thomas P. A.,
Couchman H. M. P., White S. D. M., Efstathiou G., Peacock J. A. \& Nelson A. H.
1999, ApJ, 521, 99

Ritchie B. W. \& Thomas P. A. 2000, MNRAS, 323, 743 (RT)

Rosen A. \& Bregman J. N. 1995, APJ 440, 634.

Shapiro P. R., Martel H., Villumsen, J. V. \& Owen J. M. 1996, ApJS, 103, 269.

Somerville R. S. \& Primack J. R. 1999, MNRAS, 310, 1087

Springel V. 2000, MNRAS, 312, 859

Springel V., Yoshida N., White S. D. M. 2001, New Astronomy, 6, 79 (SYW)

Springel V., Hernquist L. 2002a, MNRAS, 333, 649

Springel V., Hernquist L. 2002b, astro-ph/0206393, submitted to MNRAS

Steinmetz M. \& Mueller E. 1994, A\&A, 281L, 97

Sutherland R. S. \& Dopita M. A. 1993, ApJS, 88, 253 (SD)

Thacker R. J. \& Couchman, H. M. P. 2000, ApJ, 545, 728

Thacker R. J. \& Couchman H. M. P. 2001, APJL, 555, 17.

Theuns T., Leonard A., Efstathiou G, Pearce F. R. \& Thomas P. A. 1998, MNRAS, 301, 478

Wada K. \& Norman A. N. 2001, APJ 547, 172.

Weinberg D. H., Hernquist L. \& Katz N. 1997, ApJ, 477, 8

White S. D. M. \& Frenk C. S. 1991, ApJ, 379, 52

Wolfire M. G., Hollenbach D., McKee C. F., Tielens A. G. G. M. \& Bakes E. L. O. 1995, ApJ, 443, 152



\end{references}
\end{document}